\documentclass[%
 aip,
 amsmath,amssymb,
preprint,%
 reprint,%
]{revtex4-1}

\UseRawInputEncoding\usepackage{bm}
\usepackage{natbib}
\usepackage{graphicx}

\begin{document}
\textfloatsep 10pt

% Author Orchid ID: enter ID or remove command
\newcommand{\orcidauthorA}{0000-0002-3912-4727} % Add \orcidA{} behind the author's name

\title{Turbulence-assisted formation of bacterial cellulose}

% Authors, for the paper (add full first names)
\author{ Sung-Ha Hong, Jia Yang, Mahdi Davoodianidalik, Horst Punzmann, Michael Shats and Hua Xia}
%\author{ }%\orcidA{}}
\email{hua.xia@anu.edu.au}

\affiliation{Research School of Physics, The Australian National University, Canberra ACT 2601, Australia}

\date{\today}

\begin{abstract}

Bacterial cellulose is an important class of biomaterials which can be grown in well-controlled laboratory and industrial conditions. The cellulose structure is affected by several biological, chemical and environmental factors, including hydrodynamic flows in bacterial suspensions.  In this work, we explore the possibility of using well controlled turbulent flow to control the bacterial cellulose production. Turbulent flows leading to random motion of fluid elements may affect the structure of the extracellular polymeric matrix produced by bacteria. Here we show that two-dimensional turbulence at the air-liquid interface generates chaotic rotation at a well-defined scale and random persistent stretching of the fluid elements.  The results offer new approaches to engineering of the bacterial cellulose structure by controlling turbulence parameters.

\end{abstract}

\maketitle

\section{Introduction}
Bacterial biofilms represent the most abundant form of life on Earth \cite{Hall-Stoodley2004}. Microorganisms form extracellular polymeric matrices in a fluid environment, typically at the liquid-air, liquid-liquid and liquid-solid interfaces. Some of the biofilms, such as the bacterial cellulose (BC) \cite{Jonas1998,Campano2016}, are valuable biomaterials with great industrial and medical potential \cite{Nguyen2005,Svensson2005,KL2015}. The BC is produced by various bacteria, for example, by the genera Acetobacter, Acanthamoeba, and Achromobacter, at the liquid-air interface in a variety of forms, such as cellulose sheets \cite{Yamanaka1989}, or spherical cellulose beads (SCB) \cite{Hu2013}. The SCB is a desired cellulose form in many applications \cite{Gemeiner1989, Petersen2011,Picheth2017}. Spherical beads make perfect delivery vehicles for many functional materials, such as graphene, magnetic and conducting metal particles and other chemical compounds which can be encapsulated into BC spheres \cite{Iguchi2000,Drozd2019}. 

An attractive feature of the BC is the possibility to create cellulose with a desired fibre structure at both macroscopic and nanoscopic scales in the process of the material formation, rather than using post-production treatment methods. Along with chemical and biological processes which affect the biofilm development, hydrodynamic flows in bacterial suspensions are also capable of shaping biofilms \cite{Stone}. It has been recently demonstrated that when suspensions of \textit{E. coli} are perturbed by surface waves, patterned biofilms at the liquid-solid interface are developed \cite{Hong_SA_2020}. 

Surface waves can form ordered patterns, but they can also lead to the generation of turbulent flows at the air-liquid interfaces \cite{Kameke,Francois2013,Francois2014}. Such flows were shown to be very similar to quasi-two-dimensional (2D) turbulence which can be generated in laboratory \cite{Tabeling,Xia2008,XiaNP2011}. Turbulence greatly enhances particle dispersion \cite{Bodenschatz2010,Barkley2015,Sreenivasan2019}. Thus, it can significantly affect the growth of the extracellular polymeric matrix during the biofilm development and can change the resulting structure of the cellulose. Since the wave-driven fluid motion decays in the bulk of the liquid \cite{Falkovichbook,Kameke3D}€™, turbulent motion is the strongest in a relatively thin layer (a fraction of the wavelength) near the liquid-air interface. The BC produced at the air-liquid interfaces is particularly suitable to test this idea. Since the microorganisms forming the BC typically are not motile, flows in bacterial suspensions should not be affected by the motion of bacteria, as it occurs in suspensions of fast swimming micro-swimmers which can generate turbulence themselves \cite{Goldstein2013}. 

Turbulence is a strongly non-equilibrium state of a flow characterised by broad kinetic energy spectra and enhanced dispersion of fluid particles \cite{Turbdispersion}.  Finite-size particles in turbulence experience random walk while they are also exposed to random torques causing rotational diffusion \cite{Zeff_Nature2003}. At small scales, the rotational diffusion coefficient associated with the Brownian motion strongly decays with the increase of the particle size $d$: $D_{rot} \propto T/(\mu d^3)$ (where $\mu$ and $T$ are the fluid viscosity and its temperature) and it is negligibly small in the millimetre range of scales \cite{Berg1983}. However, in 2D turbulence, the particle dispersion is governed by the scale comparable to the agitation (forcing) scale \cite{Xia2013,Xia2014,Xia_JFM_2019}. Therefore, this scale may affect the characteristic scale of the bacterial fibre structure. In addition, turbulent flows generate stretching of the fluid elements which can also affect the formation of the bacterial cellulose.

Here we show that turbulence indeed generates chaotic rotation in a fluid. The angular velocity of turbulent rotation is the strongest at the scale just below the forcing scale of turbulence, while above this scale it decays exponentially. Such turbulent rotation can lead to a dramatic effect on the formation of the \textit{G. xylinus} BC near the liquid surface. Turbulence also effects the microscopic structure of the BC, namely, the biofilm porosity due to chaotic stretching of bacterial suspensions.

\section{Results}

\begin{figure*}
\includegraphics[width=15cm]{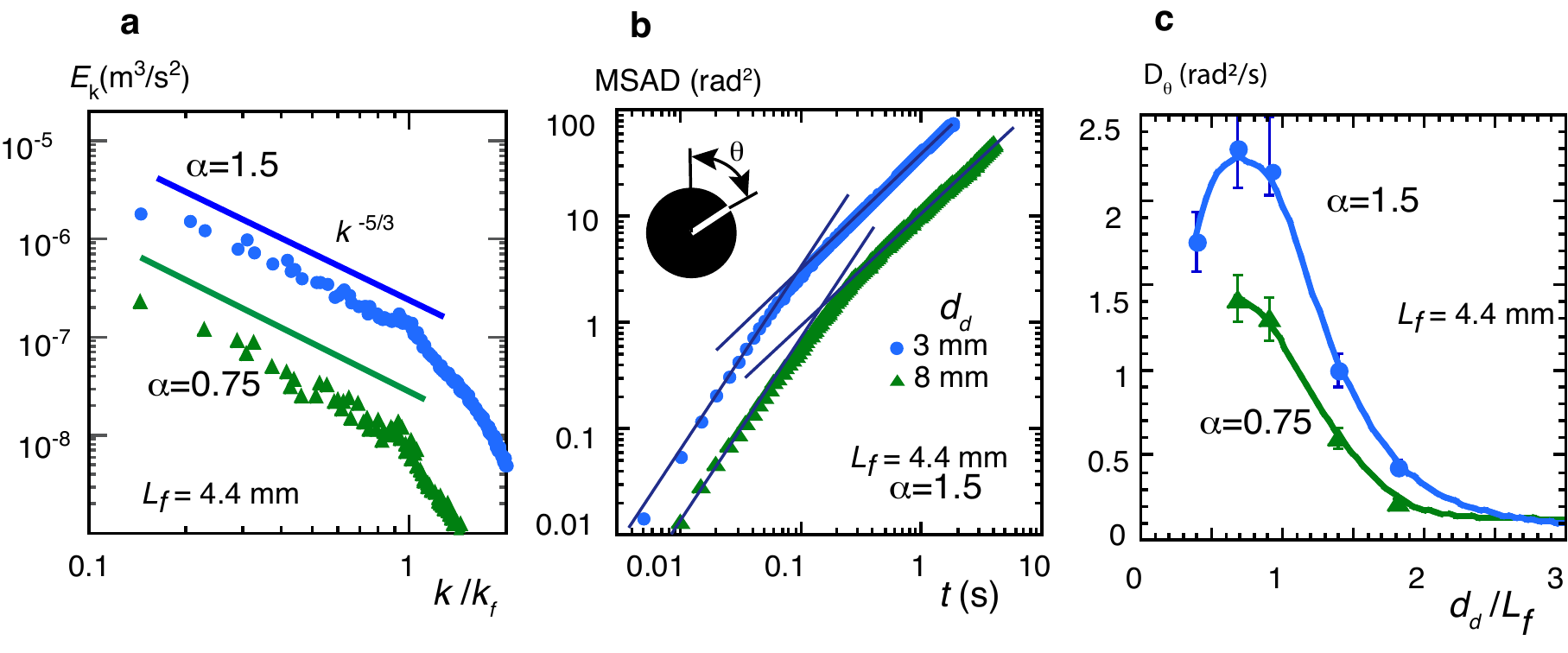}
\caption{\label{fig4} Turbulent rotation and angular displacement characteristics in 2D turbulence. (a) Kinetic energy spectra on the surface of the liquid at two levels of the vertical acceleration $\alpha=0.75$ (green triangles) and $\alpha=1.5$ (blue circles). (b) Mean squared angular displacement of the circular disks versus time: disk diameters $d_d= 3$ mm (blue circles) and 8 mm (green triangles). The forcing scale of turbulence is $L_f=4.4$ mm, supercriticality $\alpha=1.5$. At short times, MSAD$\propto t^2$, indicating the ballistic regime. At longer times, MSAD$=D_{\theta} t$, indicating diffusive regime. (c) The rotational diffusion coefficients versus the ratio  $d_d/L_f$ of the disk diameter over turbulence forcing scale. These are measured at two levels of the supercriticality $\alpha=0.75$ (green triangles) and $\alpha=1.5$ (blue circles). }
\end{figure*}

\subsection{Turbulent rotation and stretching in quasi-2D turbulence}

It is recognised that the BC formation in agitated or shaken culture methods can lead to the production of different particle sizes and various shapes (spherical, ellipsoidal, fibrous suspensions etc.) \cite{Wang2019}. The relationship between the flow conditions and the resulting shape of the BC particles however is not established. The wave-driven turbulence provides an opportunity to study the BC development under well-controlled conditions since the statistical characteristics of such flows can be accurately characterised by a few key parameters: turbulence forcing scale, and the turbulence kinetic energy which is proportional to the supercriticality $\alpha$ (see above)\cite{Xia2013,Xia_JFM_2019}. These parameters control the diffusive motion of fluid particles at the liquid surface \cite{YangPRFluids2019,Yang_PoF_2019,Xia_PoF_2019}. In addition to displacement of particles in turbulence from their initial position in a random-walk fashion, turbulent flows also lead to random rotation of finite size objects due to the generation of turbulent torque \cite{ZimmermannPRL2011}. Such torque can lead to bending of the BC fibres and to the changes in the macroscopic structure of the cellulose. In quasi-2D turbulence, it is expected that the rotation is small at the scales smaller than the forcing scale since the flow consists of coherent bundles of fluid particles \cite{FrancoisPRF2018,Xia_PoF_2019}. However, if a particle is exposed to the forces produced by several of such bundles, one can expect that turbulent torques will lead to measurable fluctuations in the particle angular velocities. The rotational diffusivity in such flows has not been studied before.

Similarly to random translation of particles, the rotational diffusivity can be characterised by the mean square angular displacement (MSAD) $\delta {\theta}^2(t)=\sum_{0}^{t} |\Delta \theta (t)|^2$, where $\Delta \theta(t)=\theta(t+\Delta t)-\theta(t)$. The MSAD can be tracked in time by following the orientational dynamics of the disks floating on the surface of the liquid (pictured in the inset of Fig. 1(b), see Methods). No mean rotation is observed, as expected in an isotropic turbulent flow. In wave-driven, quasi-2D turbulence, kinetic energy spectra are determined by the inverse energy cascade range ($E_k \propto k^{-5/3}$ at $k \leq k_f$) and the forward enstrophy cascade ($E_k \propto k^{-3}$ at $k \geq k_f$) \cite{Kraichnan1967}. Figure 1a shows the measured spectra in these experiments. The MSAD corresponding to these spectra show that turbulent rotation of the disks is described by the ballistic rotation regime at short times, and diffusive rotation at longer times, Fig. 1b:
\begin{eqnarray}\label{eq3}
\langle \delta {\theta}^2 \rangle \approx V_{\theta}^2 t^2,    \;    \;     & t \ll T_{\theta}  \\
\langle \delta {\theta}^2 \rangle \approx D_{\theta} t,  \;   \;    & t \gg T_{\theta}
\end{eqnarray}
\noindent Here $T_{\theta}$ is the autocorrelation time of the angular velocity $T_{\theta}=1/\langle \tilde{\omega}^2 \rangle \int_0^\infty \langle {\omega}(t_0+t) {\omega}(t_0) \rangle dt$, where ${\omega}=d \theta/dt$ is the angular velocity of the disks, and $\langle \tilde{\omega}^2 \rangle $ is its variance.

For the cellulose formation occurring on the time scales of hours, we are interested in the long-time behaviour. We measure the rotational diffusion coefficient $D_{\theta}$ at different ratios of the disk diameter $d_d$ to the forcing scale $L_f$. This coefficient, shown in Fig. 1c, is peaked at $d_d/L_f <1$ and it decays exponentially at larger $d_d/L_f $. This behaviour is also observed in a different type of quasi-2D turbulence, which is driven electromagnetically. The rate of diffusion is proportional to the turbulence kinetic energy: the increase in the variance of the turbulent velocity fluctuations $\langle u^2 \rangle$ leads to the increase in $D_{\theta}$, Fig. 1c. Such turbulent torques may lead to random bending of the BC fibres resulting in the formation of the BC beads whose size is correlated with the turbulence forcing scale $L_f$. 

\begin{figure*}
\includegraphics[width=12cm]{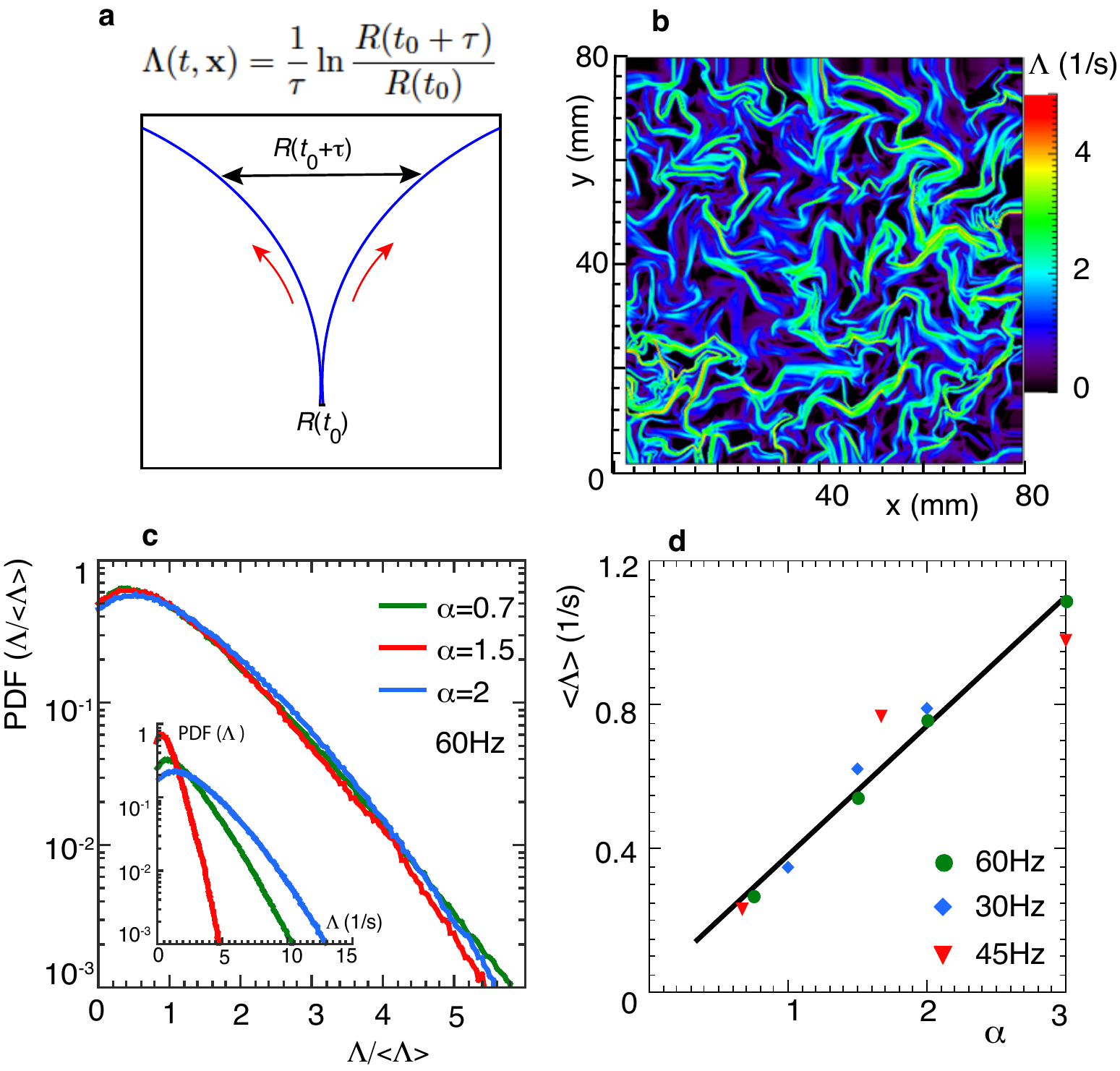}
\caption{\label{fig5} Turbulent stretching of fluid parcels. (a) Schematic of the fluid particle separation. (b) Finite-time Lyapunov exponents $\Lambda$ on the turbulent surface computed over time $\tau = 0.6 T_L$. (c) Probability density functions of $\Lambda/ \langle \Lambda \rangle$ (where $\langle \Lambda \rangle$ is the spatially averaged quantity) at three different vertical accelerations $\alpha=$0.7, 1.5 and 2 and the frequency of $f=60$ Hz. The inset shows the PDFs of non-normalised $\Lambda$. (d) Spatially averaged value of the Lyapunov exponents versus the supercriticality parameter $\alpha$ at different excitation frequencies: $f=30$ Hz (blue diamonds), $f=45$ Hz (red triangles), and $f=60$ Hz (green circles).}
\end{figure*}

To understand how turbulence can affect the microscopic structure of the BC, we study the stretching of the fluid parcels \cite{Aref1984,Ottino1989}. Two adjacent fluid particles remain close to each other for a long time as they move within a coherent bundle. Occasionally these bundles split leading to the stretching of the fluid parcels. Extreme stretching events have been found in numerical simulations of 3D turbulence \cite{Scatamacchia2012,Goto2004,Rossi2006}. However turbulent stretching has not been studied in 2D turbulence and it is not clear how the probability of stretching events depends on the turbulence parameters.

Stretching can be characterised by the relative separation of a pair of initially adjacent particles at times $t_0$ and $(t_0+\tau)$: $R(t_0+\tau) /R(t_0)$, (Fig. 2a). The logarithm of stretching normalised by time $\tau$ gives the value of the finite-time Lyapunov exponent $\Lambda(t,\mathbf{x}) = (1/\tau) \ln [R(t_0+\tau)/R(t_0)]$ which characterises the rate of local deformation of an infinitesimal fluid element \cite{Boffetta_2001}. Here $\mathbf{x}$ is the initial position of the pair of particles.

Trajectories of the fluid particles are determined using simulated trajectories derived from experimentally measured velocity fields (see Methods). The finite-time Lyapunov exponent field $\Lambda (\tau,\textbf{x})$ gives the spatial distribution of the stretching rates for a particular integration time $\tau$. Of a particular interest here, is the probability of strong stretching events occurring on the time scale comparable to the Lagrangian velocity correlation time $T_L=1/ \langle \tilde{u}^2 \rangle \int_0^\infty \langle \textbf{u}(t_0+t) \textbf{u}(t_0) \rangle dt$, where $\textbf{u}$ is the fluid velocity along the Lagrangian trajectory, $\langle \tilde{u}^2 \rangle $ is its variance, and the angular brackets denote the ensemble averaging. Such short-time Lyapunov exponents have not been studied in 2D turbulence before.

Figure 2b shows a snapshot of the stretching field computed for $\tau=0.6T_L$. The ridges of high values of $\Lambda$ constantly move and reconnect (see Supplementary Video 1), such that over time, all points on the liquid surface experience many stretching events.  The probability density functions of $\Lambda$ are illustrated in the inset of Fig. 2c for three values of vertical accelerations which correspond to the supercriticality parameter $\alpha$ from 0.7 to 2. However, if $\Lambda$ is normalised by its spatially averaged value $\langle \Lambda \rangle$, the PDFs measured at different energies collapse onto a single function, Fig. 5c. The PDF of the Lyapunov exponent can be fitted with a Weibull distribution, $f(x)=(k/\beta) (x/\beta)^{(k-1)} e^{- (x/\beta)^k}$, where $k=1.35$  is the shape parameter and $\beta=1.3$ is the scale parameter. The Weibull distribution is related to a number of other probability distributions, in particular, it interpolates between the exponential distribution ($k = 1$) and the Rayleigh distribution ($k = 2$). 

The collapse of the PDFs of the average-normalised finite-time Lyapunov exponents to this universal form is due to the fact that the spatially averaged Lyapunov exponents $\langle \Lambda \rangle$ increase approximately linearly with the increase in the supercriticality $\alpha$ of the wave-driven turbulence, Fig. 2d.

\section{Conclusions}

Turbulence generated by parametrically excited waves in vertically shaken containers (Faraday waves) affects the formation of bacterial cellulose by \textit{Gluconacetobacter xylinus} microorganisms.  It is hypothesised that the turbulence-induced torque manifested as chaotic rotation of fluid elements, leads to the bending of the cellulose fibres. Such a rotation is studied by analysing the rotation of the circular disks at the liquid surface. It is found that the maximum in the rotational diffusion coefficient corresponds to the ratio of the disk diameter over the forcing scale of $d_d/L_f \approx 0.7$. For larger disks (or smaller $L_f$), $D_{\theta}$ decreases exponentially. Turbulent rotation results from the interaction between meandering coherent bundles of fluid particles\cite{Xia2013,FrancoisPRF2018,Xia_PoF_2019}  whose widths is about $L_f$, therefore the fibre clusters whose size is much less than $L_f$ do not experience strong torque. 

Summarising, we investigate two fundamental processes in wave-driven turbulence, turbulent rotation and stretching. It is shown that turbulence can be used as a tool to produce biofilms with remarkably different macro- and microscopic structures. The results offer new methods which allow to engineer bacterial cellulose properties (e.g. porosity, water holding capacity, etc.) by employing well-controlled wave-driven turbulence. 

\section{Methods}

\textit{Turbulent flow generation}

Faraday waves are generated on the surface of a liquid which is vertically vibrated at the frequency $f_s$ with sinusoidal acceleration $a$ above a parametric excitation threshold ($a_{th}$) \cite{Faraday}, at a supercriticality parameter $\alpha=(a-a_{th})/a_{th} \ge 1$. The flow generated on the surface at high $\alpha$ exactly reproduces the statistics of 2D turbulence \cite{Kameke,Francois2013,Xia2014}. In the experiments, turbulence kinetic energy is proportional to the supercriticality parameter \cite{Xia_JFM_2019}. The kinetic energy spectrum has a form $E_k=C\epsilon^{2/3}k^{-5/3}$ in the inverse energy cascade range. The energy injection rate (equal to the dissipation rate)  $\epsilon=(E(k)/C)^{3/2}k^{5/3}$ where $C=6$ is the Kolmogorov constant and it increases with the increase in $\alpha$ as $\epsilon \sim \alpha^3$. The forcing scale of the flow is half the wavelength of the Faraday waves, $L_f=1/2 \lambda$. The experiments are conducted at $f_s=(30 - 150)$ Hz, $\alpha=(0.1-2)$, and $L_f=(2 - 13)$ mm.

Floating disks of various sizes $d_d$ (normalised size $d_d/L_f$ ranging from 0.4 to 7) are used to study the rotational torque in the flow. A white stripe is marked on each disk to track its orientation $\theta$ ($\theta \in [-90^{\circ},+90^{\circ}]$) . The angular velocities $V_\theta$ are calculated based on the angular displacements between two adjacent image frames and averaged over three points in time. The statistics of the disk rotation is captured along the disk trajectories in turbulence. To minimise the disturbance to the flow, only one disk on the surface is used in each measurement. More than 600 independent realisations (movies) are recorded in every experiment to achieve statistical convergence.

The velocity field is measured using a particle image velocimetry technique. To determine the FTLE field from the measured velocity fields, we simulate the particle trajectories by integrating the velocity field from $t=t_0$ and determine how much initially adjacent particles are separated during the finite integration time $\tau$ using a 4th order Runge-Kutta integration method.  The FTLE field gives the spatial distribution of the stretching rate within a flow for the integration time $\tau$. The mean FTLE is obtained by spatially averaging over the fluid surface characterizing the mean stretching rate of the fluid parcels.

\acknowledgments{
This work was supported by the Australian Research Council Discovery Projects and Linkage Projects funding schemes (DP160100863, DP190100406 and LP160100477). H.X. acknowledges support from the Australian Research Council's Future Fellowship (FT140100067). The authors acknowledge the technical assistance of Centre for Advanced Microscopy (ANU) and the Australian National Fabrication Facility (ACT node).

Author contributions:

H.X and M.S designed the project. S.H.H, M. D and J. Y conducted the rotational diffusion experiments. H.P. designed the experimental setup. J.Y. and H.X. analysed the experimental data. H.X and M.S wrote the paper. All authors reviewed the paper. 
}

\end{document}